# Development of Multi-Layer Fabrication Process for SFQ Large Scale Integrated Digital Circuits

Liliang Ying[1,2], Xue Zhang[1,2], Minghui Niu[1,2], Jie Ren[*1,2,3], Wei Peng[1,2,3], Masaaki Meazawa[1,2], and Zhen Wang[1,2,3]

[1]Shanghai Institute of Microsystem and Information Technology (SIMIT), Chinese Academy of Sciences (CAS), Shanghai 200050, China

[2]CAS Center for Excellence in Superconducting Electronics (CENSE), Shanghai 200050, China

[3]Univerisity of Chinese Academy of Science, Beijing 100049, China

***Abstract*—We have developed a fabrication technology for the development of large-scale superconducting integrated circuits with Nb-based Josephson junctions. The standard fabrication process with 10 mask levels uses four metal layers including 3 Nb superconducting layers and a Mo resistor layer. The influence of deposition parameters on film stress, electrical properties, and surface roughness were studied systematically. High quality Nb, Al, Mo, and $SiO_2$ films were successfully deposited for the subsequent fabrication of circuits. The circuit fabrication started with the fabrication of Mo resistors with a target sheet resistance $R_{sh}$ of 2 Ω, followed by the deposition of Nb/Al-$AlO_x$/Nb trilayer Josephson-junction. The target critical current density $J_c$ was set at 6 kA/cm$^2$. The thicknesses and etch depths of the films were monitored during fabrication with on-wafer process-control-monitor (PCM) patterns for all the wafers. The fabrication process has been evaluated by electrical testing of specially designed PCM circuits. Circuit parameters such as $J_c$ are considered for the evaluation purposes. Small-scale circuits such as our standard library cells have been successfully fabricated and tested, confirming the usefulness of our fabrication technology for superconducting integrated circuits.**

## I. Introduction

Rapid single flux quantum (RSFQ) superconductor electronics (SCE) based on Nb/Al-$AlO_x$/Nb Josephson junctions (JJs) have received much attention due to their very high processing speeds and low power dissipations. Nb/Al-$AlO_x$/Nb Josephson junctions are finding applications for high-performance computing. Toward real-world applications of the large-scale superconducting digital circuits, numerous efforts have been made to improve fabrication technologies, for example, by MIT Lincoln Laboratory (MIT LL) [1-3] and HYPRES [4-6] in the USA, National Institute of Advanced Industrial Science and Technology (AIST) in Japan [7-9], Leibniz Institute of Photonic Technology (IPHT) in Germany [10].

The fabrication was performed in the superconducting electronics facility (SELF) of SIMIT. This work was supported by the Strategic Priority Research Program of Chinese Academy of Sciences (Grant No. XDA18000000), Shanghai Science and Technology Committee (Grant No. 17JC1401100), and Chinese Academy of Sciences Key Technology Talent Program.

(Corresponding author: Jie Ren：jieren@mail.sim.ac.cn)

JJs are key elements of high-performance RSFQ circuits. Higher-critical-current-density ($J_c$) and smaller-area JJs improve both operating speeds and integration densities of the superconducting circuits, which is a strong requirement for the fabrication technology. In other words, a reliable fabrication process that can stably produce superconducting LSI chips with a large number of high-$J_c$, small-area JJs is essential to the high-performance SCE applications such as next-generation extremely-low-power supercomputers.

Our method toward superconducting LSIs with high-$J_c$, small-area JJs is deposition of Nb/Al-$AlO_x$/Nb JJ as early stage of the wafer fabrication process. This technique has several advantages: improving the uniformity and repeatability of JJ parameters, increasing resolution and uniformity of photolithography, narrowing distribution of JJ critical current, and mitigating the impact of stress on JJ quality. On the other hand, in most of the fabrication processes of Nb/Al-$AlO_x$/Nb JJs, the junction barrier Al-$AlO_x$ layer is etched by ion beam etching (IBE) [9] or wet etching [11]. However, during the ion beam etching, sidewalls are sometimes formed at the edge of etched profiles due to the re-deposition effect. These sidewalls may create additional conductive paths in the insulation layer to cause a current leakage in the junctions. On the other hand, conventional manual wet etching involving chemical reaction degrades the process stability due to difficulties in controlling etch parameters such as temperature and time. In this work, we opted the Al-$AlO_x$ removal by a photoresist developer in a clean track system. Such a method using automatic equipment is more stable and reliable compared to the manual wet etching and also avoids the sidewall formation caused by IBE. In short, our proposed process maintains high yield at low cost by reducing number of fabrication steps and avoiding sidewalls during the Al-$AlO_x$ etching.

In this paper, we present our standard fabrication process for Nb-based large-scale superconducting integrated circuits. The process consists of with 10 mask levels including four metal layers of three Nb superconducting layers and one Mo resistor layer. Thermally-oxidized 4-inch Si wafers were used for substrates. The targeted $J_c$ for Nb/Al-$AlO_x$/Nb JJs is 6 kA/cm$^2$ and minimum JJ size is ϕ 1.4 μm. The sheet resistance of the Mo resistor layer was 2 Ω. The deposition parameters for Nb/Al-$AlO_x$/Nb trilayer films were optimized for high flatness, high $T_c$, and low stress [12]. The process parameters for elementary process steps such as lithography and etching are also carefully adjusted and optimized. The process parameters were systematically monitored with our process control monitor (PCM) patterns [13], making routine feedback for the process improvement. We eventually fabricated small-scale RSFQ circuits including our library cells [14] which were successfully tested.

## II. Fabrication Process

### A. *Nb/Al-AlO$_x$/Nb trilayer deposition*

We investigated the deposition parameters in detail for single-layer Nb and Al, and Nb/Al-AlO$_x$/Nb trilayer films, which were deposited with a load-locked multi-chamber sputtering system. Nb films were deposited by DC magnetron sputtering at the sample-holder temperature of room temperature. The surface morphology of the Nb film was investigated using an AFM with a scan area of 1 μm × 1 μm. Fig. 1 shows the root mean square (RMS) roughness of films deposited with different sputtering currents and Ar pressures. It was revealed that the RMS roughness decreased as the applied sputtering current increases and roughness increased as Ar pressure was increased. Our previous studies showed that the roughness of low Nb content films should preferably be less than 2 nm[12]. To avoid an effect of unexpected fluctuations in the process, we chose the parameters that yielded a roughness less than 1.5 nm.

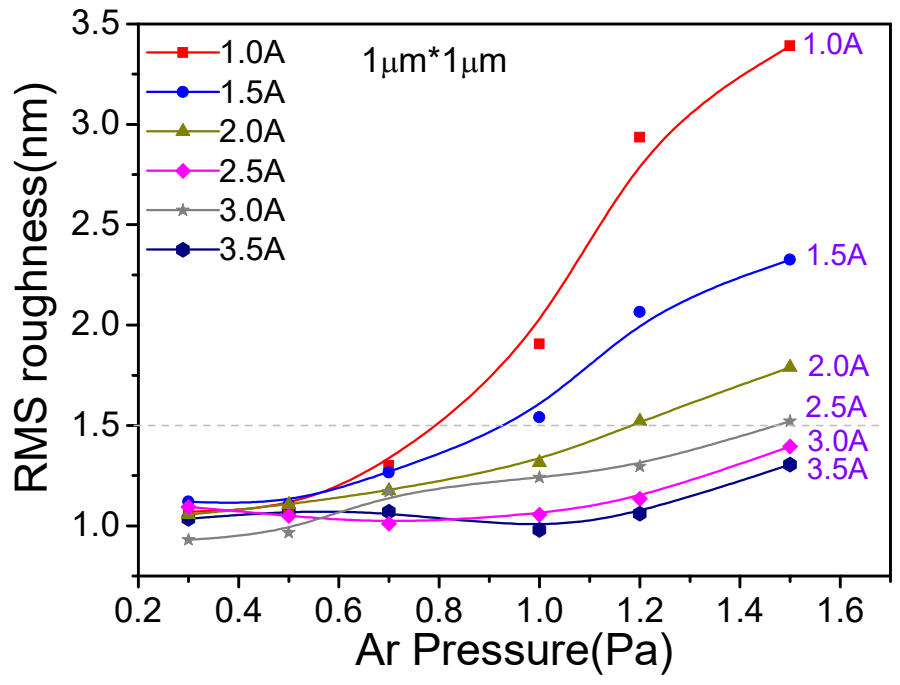

Fig. 1. RMS roughness of surfaces of Nb films as a function of Ar pressure at different sputtering currents ranging from 1.0 A to 3.5 A. The observed area is 1 μm x 1 μm.

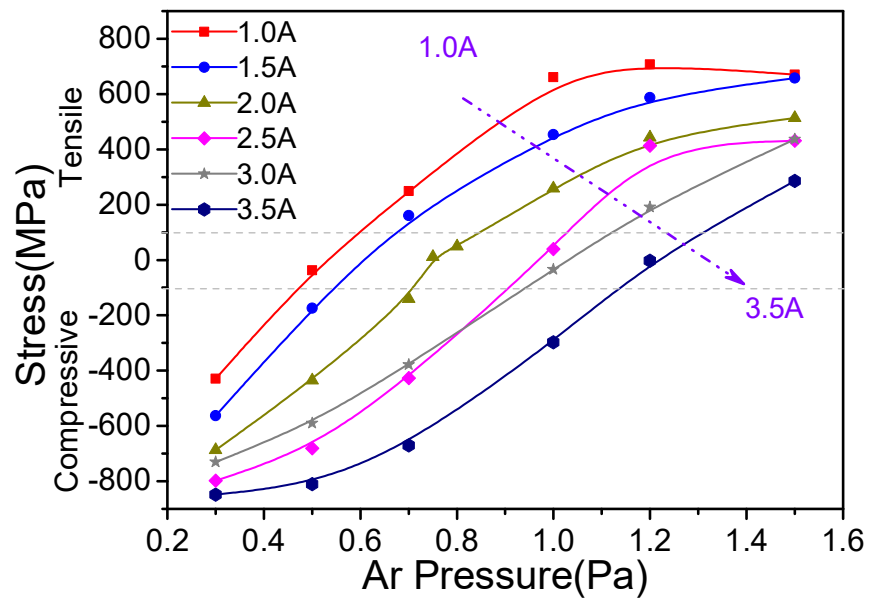

Fig. 2. Stress in Nb films as a function of the Ar pressure $P_{Ar}$ at different sputtering currents ranging from 1.0 A to 3.5 A.

The stress of 150-nm-thick Nb films was measured using a laser-based film stress measurement system (Frontier Semiconductor FSM 128). The results are plotted in Fig. 2 as a function of the Ar pressure $P_{Ar}$ for different sputtering currents. The stress monotonically increases with increasing $P_{Ar}$ in the range of 0.3 Pa to 1.5 Pa. The stress also increases as the sputtering current increases. The optimal stress target of Nb film is between -100 MPa and 100 MPa, which can be achieved adjusting the current and $P_{Ar}$.

Nb films deposited at different sample-holder temperatures was electrically characterized using a Quantum Design PPMS system by a standard four-point measurement. The superconducting transition temperature $T_c$ and residual resistance ratio RRR = $R_{300K}/R_{9.5K}$ are plotted in Fig. 3, where $R_{300K}$ and $R_{9.5K}$ are resistance at 300 K and 9.5 K, respectively. Our criteria for the lowest $T_c$ and RRR were fixed at 9.18 K and 5, respectively. Most of the films met the criteria. Taking all the measured data into consideration, we determined the standard parameters for the Nb film deposition to be the sputtering current of 2.0 A and Ar pressure of 0.7 Pa.

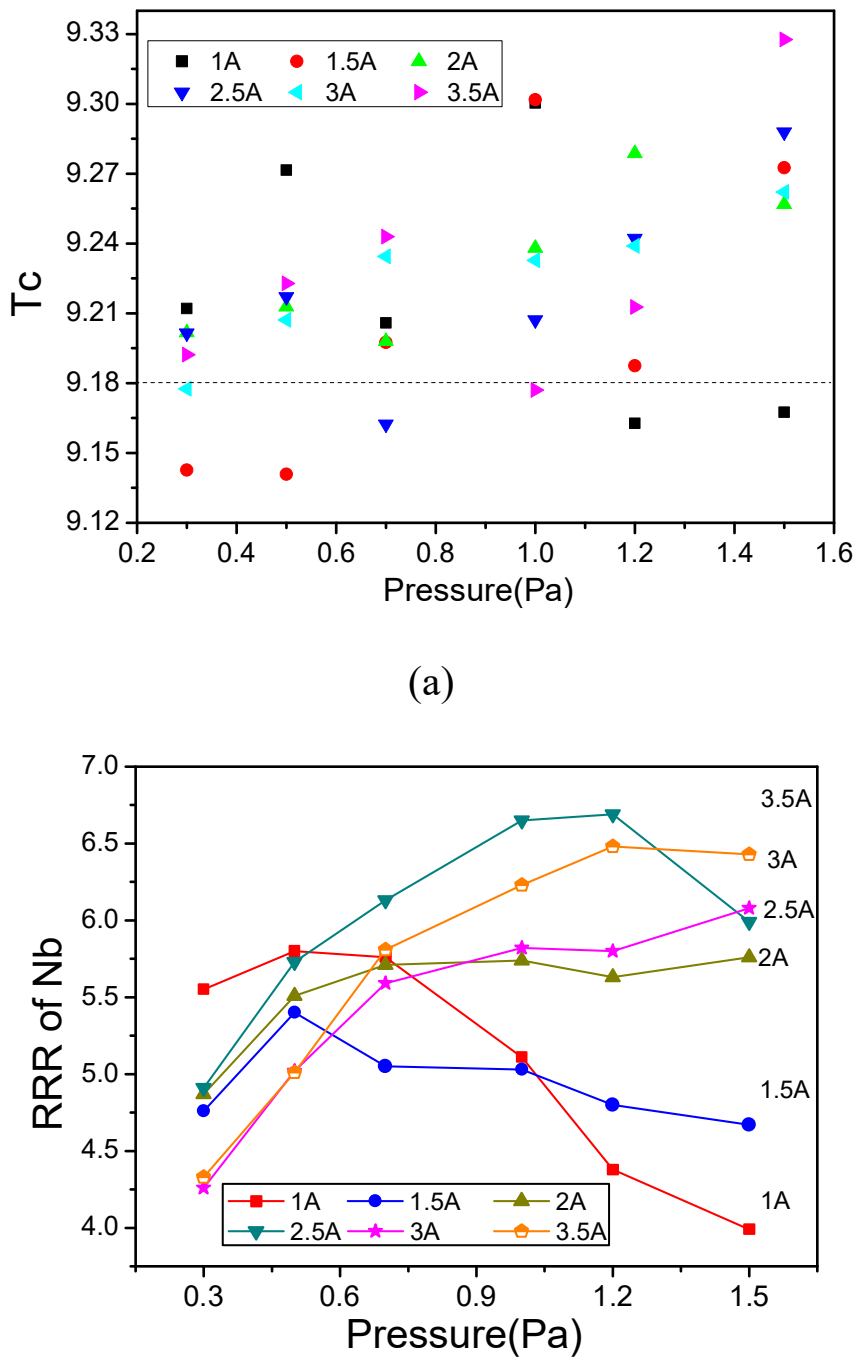

(a)

(b)

Fig. 3. Electrical characteristics of Nb films: (a) superconducting transition temperature $T_c$ and (b) residual resistance ratio RRR.

Al films were deposited by the DC magnetron sputtering system in a different chamber from the Nb deposition chamber. Fig. 4 shows RRR of the 100-nm-thick Al films measured using the Quantum Design PPMS system. To obtain Al films with RRR > 5, we determined the standard parameters for the Al deposition to be the sputtering current of 0.5 A and Ar pressure of 0.5 Pa at which RRR dependence on the Ar pressure is weak.

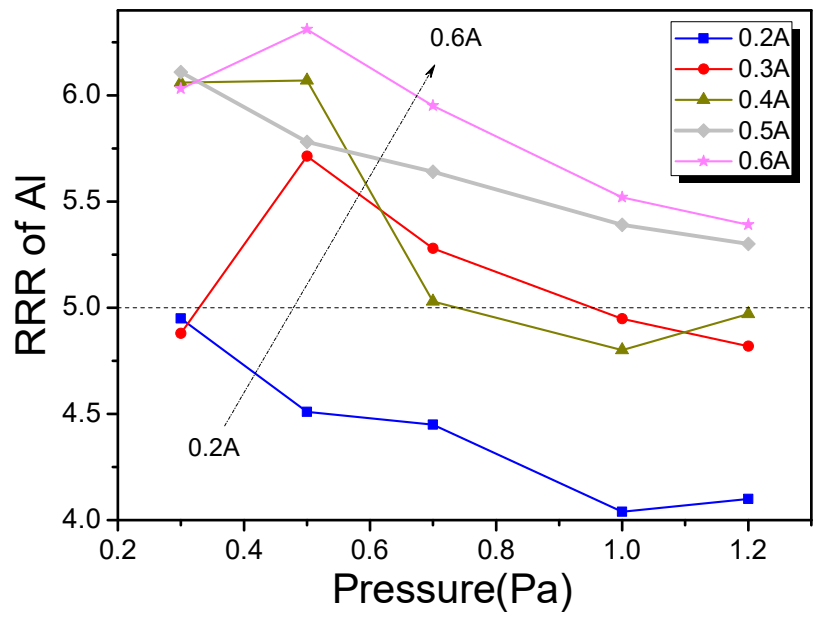

Fig. 4. RRR of Al films as a function of Ar pressure at different sputtering currents.

### B. *Circuit fabrication*

The circuit fabrication process includes a resistor layer, a trilayer, wiring, a ground plane at the top, and a contact pad metalization. Fig. 5 shows a cross-section of a device fabricated by our standard process. Table 1 summarizes the parameters of the layers, the mask levels, and minimum feature size (on the mask). The substrate was a 4-inch silicon wafer with 300-nm-thick thermal oxide. A standard chip size was 5.2 mm x 5.2 mm. Mo resistor layer with sheet resistance of 2.0 Ω was used for junction shunting and circuit biasing. The critical current density $J_c$ of the junction was fixed at 6 kA/cm$^2$ which was controlled by an $O_2$ exposure $E$, a product of $O_2$ partial pressure and time for Al oxidation. Lithography for all the layers was performed using an *i*-line (365 nm) 5-to-1 stepper.

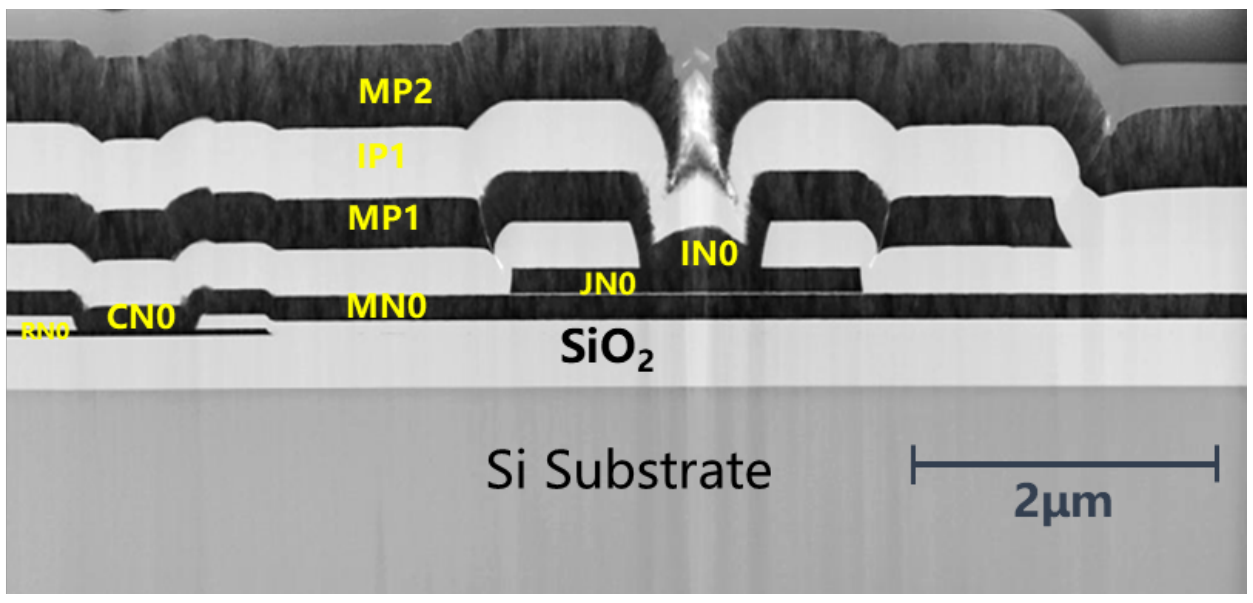

Fig. 5. Cross-section of a device fabricated by SIMIT standard fabrication process.

The process started with the deposition of a 40-nm Mo film (RN0) which was then patterned by a reactive ion etching (RIE) with an endpoint detector using $SF_6$ mixed with $O_2$ gas at 30 mTorr. A 100-nm $SiO_2$ film (CN0) was deposited using plasma-enhanced chemical vapor deposition (PECVD). Vias in CN0 layer were fabricated using RIE with $CHF_3$ at 2 Pa. Then Nb/Al-$AlO_x$/Nb trilayer film was fabricated in the multi-chamber DC magnetron sputtering system. First, a Nb film with a 150-nm thickness for the base electrode (MN0) was sputtered at $P_{Ar}$ = 0.7 Pa and a sputtering current of 2.0 A, and a 12-nm-thick Al layer was then deposited at $P_{Ar}$ = 0.5 Pa and a sputtering current of 0.5 A. An $AlO_x$ tunnel barrier (AN0) was formed by Al oxidation in the load-lock chamber at 4 Pa of a mixed $O_2$ ambient (15% $O_2$ mixed with 85% Ar) for 30 to 40 minutes. Another 150-nm Nb film for the counter electrode (JN0) was deposited using the same conditions as the first Nb film. The Nb films of both the base and counter electrodes were patterned and etched by inductively coupled plasma RIE (ICP-RIE) with an endpoint detector. The $AlO_x$ tunnel barrier (AN0) was removed by wet etching with a photoresist developer in an auto-coating/developing system. After the trilayer patterning, a 250-nm $SiO_2$ was deposited using PECVD, and vias in IN0 were made by RIE with the same parameters as those for the CN0 vias. 300 nm of Nb wiring (MP1) was deposited and then patterned by ICP-RIE with an endpoint detector. Then a 400-nm $SiO_2$ layer (IP1) was deposited using PECVD and vias in IP1 were fabricated with the same parameters as those for CN0 and IN0 vias. A 500-nm Nb ground plane (MP2) was deposited and patterned. Finally, a 100-nm Au contact pad (PP2) with an 8-nm Ti adhesion layer was deposited using electron-beam evaporation and patterned by a lift-off technique.

The layer thicknesses and etched depths were measured with the on-wafer PCM patterns during the wafer fabrication process. The alignment shifts and resolutions for the lithography processes were also monitored for all the layers. Details of the during-fabrication PCM evaluations are presented elsewhere [13].

TABLE1. LAYER PARAMETER AND MASK INFORMATION

| **Layer Name** | **Material** | | **Thickness (nm)** | **Min. size (μm)** | **Min. space (μm)** |
|---|---|---|---|---|---|
| RN0 | Mo | Resistor | 40 ± 6 | 1.6 | 1.0 |
| CN0 | $SiO_2$ | Insulation | 100 ± 10 | 1.0 | 1.0 |
| MN0 | Nb | Base electrode | 150 ± 5 | 1.6 | 1.5 |
| JN0 | Nb | Counter electrode | 150 ± 5 | 1.4 | 1.0 |
| IN0 | $SiO_2$ | Insulation | 250 ± 20 | 0.7 | 1.0 |
| MP1 | Nb | Superconductor | 300 ± 15 | 2.0 | 1.0 |
| IP1 | $SiO_2$ | Insulation | 400 ± 20 | 1.6 | 1.0 |
| MP2 | Nb | Superconductor | 500 ± 20 | 2.6 | 2.0 |
| PP2 | Ti/Au | Contact pad | 100 ± 5 | 3.0 | 3.0 |

## III. MEASUREMENT AND CHARACTERIZATION

The fabrication process has been evaluated by electrical testing using specially-designed PCM circuits[13]. Circuit parameters such as $J_c$ and $R_{sh}$ as well as the probabilities of the defects were chosen as evaluation criteria. Feedback of the PCM results has been made routinely to improve the fabrication technology.

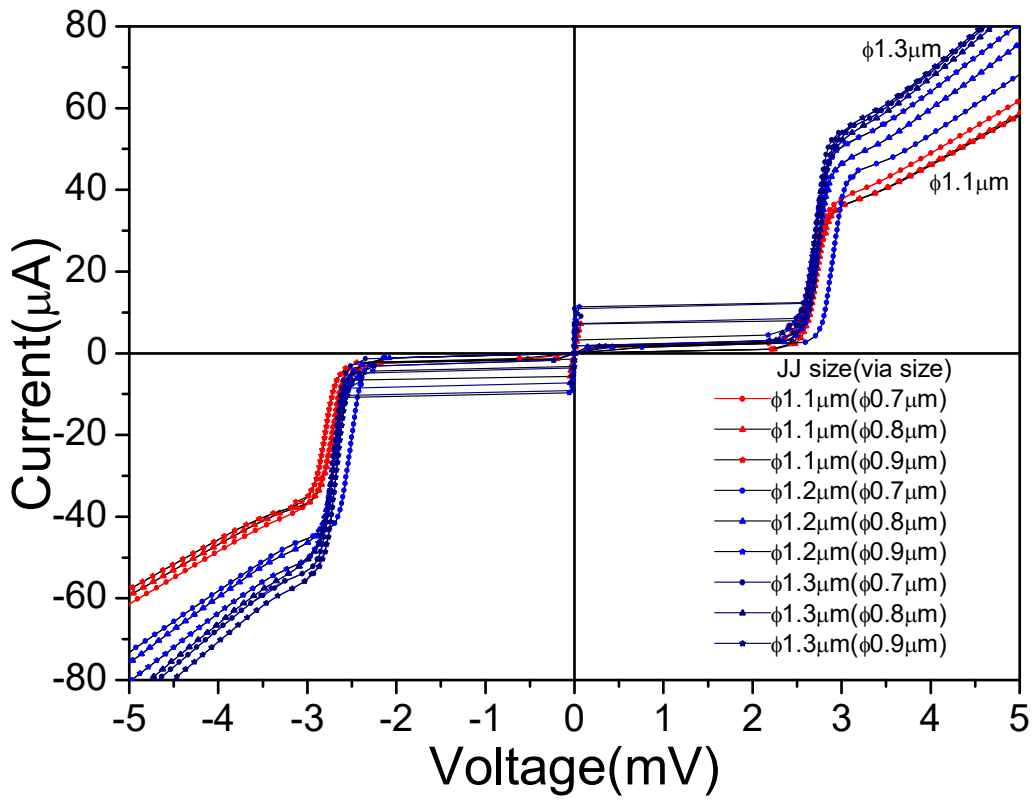

Fig. 6. Typical I-V characteristics for Nb/Al-AlO$_x$/Nb junctions fabricated by our standard process.

Parameters related to the junction quality were extracted from current voltage curves (IV curves) of un-shunted JJs. We first investigated the IV curve of a single circular un-shunted junction. Fig. 6 shows the typical I-V characteristics measured at 4.2 K for Nb/Al-AlO$_x$/Nb junctions with different sizes ranging from 1.1 to 1.3 μm in diameter. To examine the controllability of current density $J_c$, whose values were systematically determined with our PCM junctions [13], we investigated the dependence of $J_c$ on $O_2$ exposure $E$. Fig. 7 shows the $J_c$ as a function of $O_2$ exposure $E$. $J_c$ can be empirically fitted to $J_c \propto E^{-\alpha}$ [15]. The obtained $\alpha$ values change from 0.46 in the low-$J_c$ region to 2.2 in the high-$J_c$ region, which is qualitatively in a good agreement with the previously reported results [15].

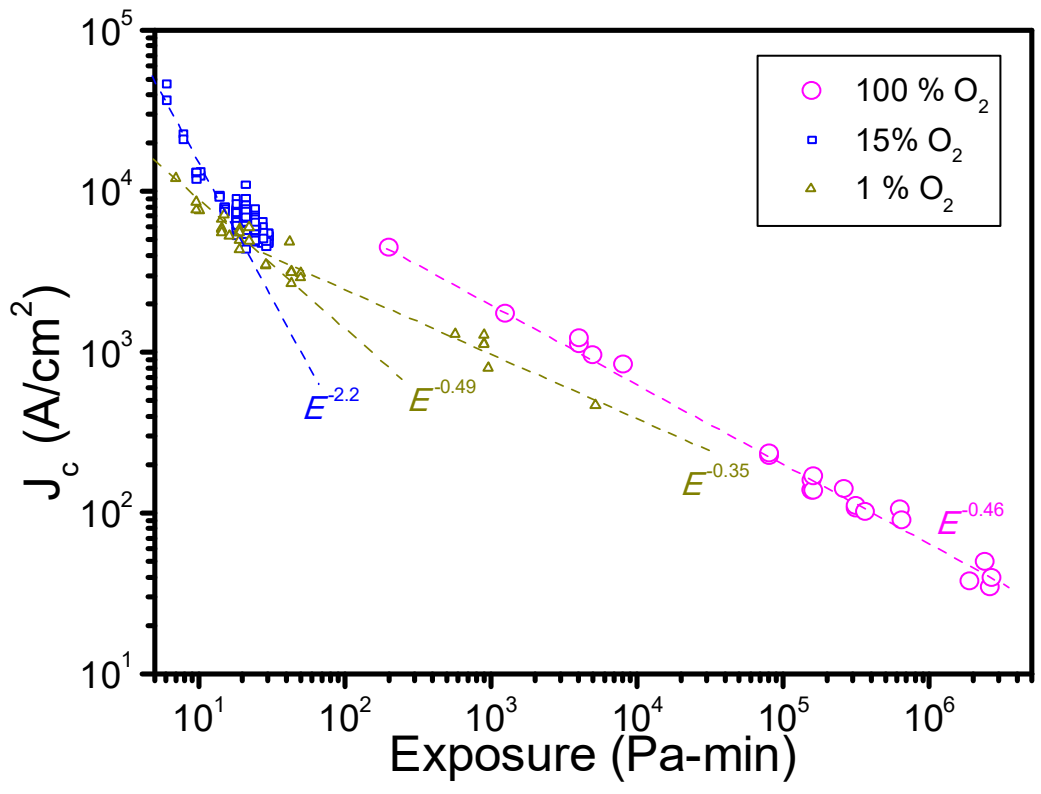

Fig.7. Dependence of $J_c$ on $O_2$ exposure.

Small-scale circuits were designed and fabricated by our standard process described above. Fig. 8 shows a microphotograph of an AND gate in our standard cell library [14] which was successfully tested at low frequencies as shown in Fig. 9. The experimental operating margins for the bias current were -35% to +40% for the AND gate.

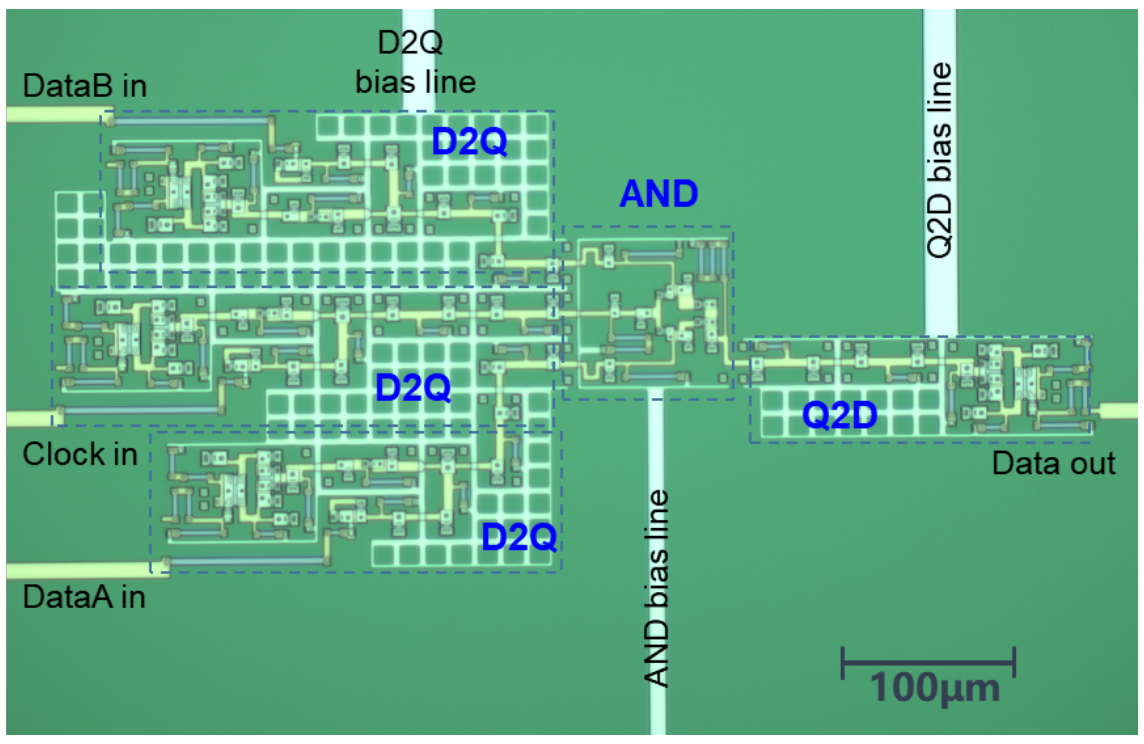

Fig. 8. Photograph of the fabricated cell.

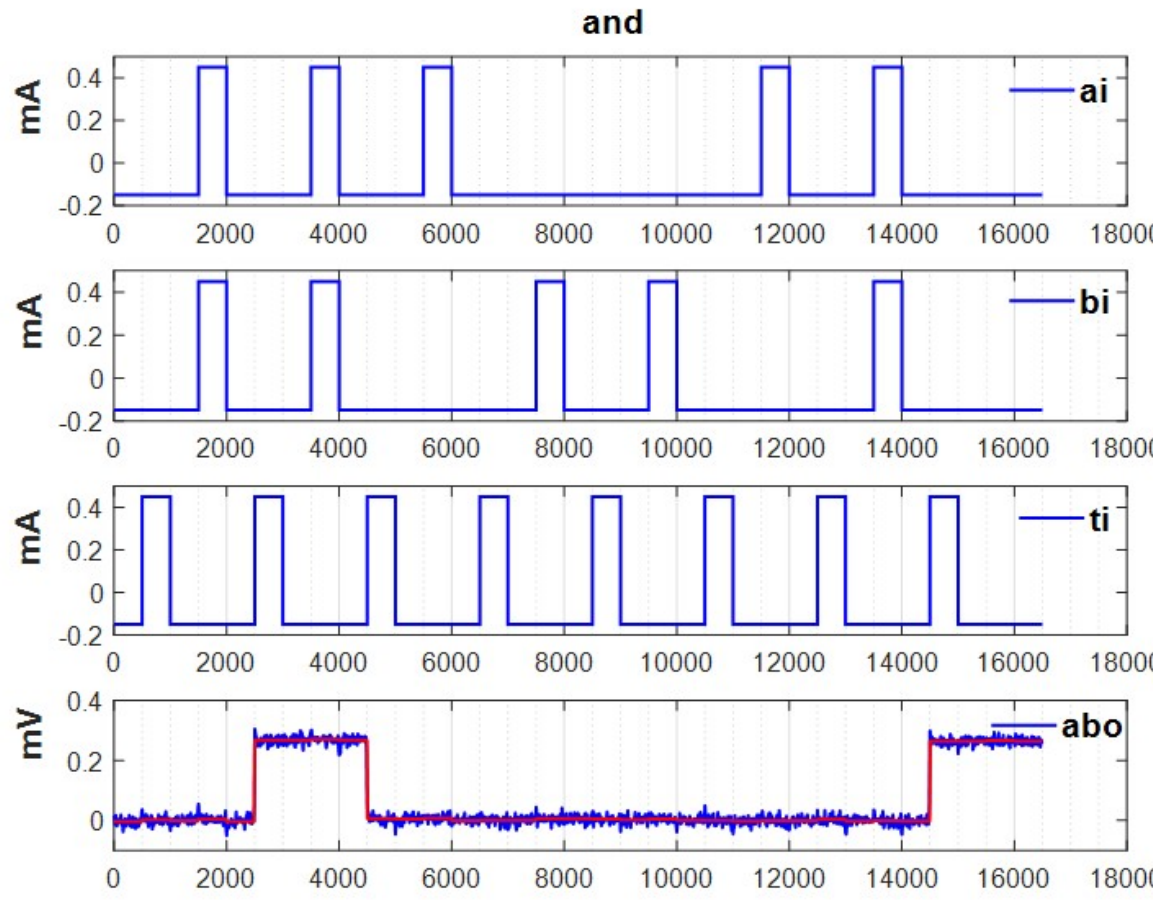

Fig. 9. Operation of an AND cell in our standard cell library.

## IV. Conclusion

We have developed our standard fabrication process with Nb/Al-AlO$_x$/Nb Josephson junctions for superconducting integrated circuits. The process includes a Nb/Al-AlO$_x$/Nb trilayer deposited in early stage of the fabrication. Al-AlO$_x$ junction barrier layers have been removed by a photoresist developer in an auto-coating/developing system. Such a method using an automatic equipment is more stable and reliable than manual wet etching, and avoids the sidewall caused by ion beam etching (IBE). We have achieved the high controllability of $J_c$. Small-scale circuits have been fabricated by the standard process and successfully tested at low frequencies.

## References

[1] S.K. Tolpygo, V. Bolkhovsky, R.Rastogiet al.,"Advanced Fabrication Processes for Superconductor Electronics: Current Status and New Developments", *IEEE Trans. Appl. Supercond.*, **vol. 29, No. 5**, 1102513, August 2019;

[2] S.K. Tolpygo, V. Bolkhovsky, T.J. Weir et al., "Advanced fabrication processes for superconducting very large scale integrated circuits,"*IEEE Trans. Appl. Supercond.*, **vol. 26, No. 3**, 1100110, April 2016;

[3] S. K. Tolpygo, V. Bolkhovsky, T. J. Weir, L.M. Johnson, M. A. Gouker, W. D. Oliver, "Fabrication process and properties of fully-planarized

deep-submicron Nb/Al-AlO$_x$/Nb Josephson junctions for VLSI circuits," *IEEE Trans. Appl. Supercond.*, **vol. 25, no. 3**, 1101312, Jun. 2015

[4] S.K. Tolpygo, D. Yohannes, R.T. Hunt et al., "20kA/cm$^2$ process development for superconducting integrated circuits with 80 GHz clock frequency," *IEEE Trans. Appl. Supercond.*, **vol. 17, No. 2**, 946-951, June 2007.

[5] D. Yohannes, S. Sarwana, S. K. Tolpygo, A. Sahu, Y. A. Polyakov, and V. K. Semenov, "Characterization of HYPRES' 4.5kA/cm$^2$ and 8 kA /cm$^2$ Nb/AlOx/Nb fabrication process," *IEEE Trans. Appl. Supercond.,* **vol. 15**, pp. 90–93, 2005.

[6] D. Yohannes, A. Kirichenko, S. Sarwana, and S. K. Tolpygo, "Parametric testing of HYPRES superconducting integrated circuit fabrication processes," *IEEE Trans. Appl. Supercond.*, **vol. 17, No. 2,** pp. 181-186, June 2007.

[7] S. Nagasawa, K. Hinode, T. Satoh, et al., "Nb 9-layer fabrication process for superconducting large scale SFQ circuits and its process evaluation," *IEICE Trans. Electron.*, **vol. E97-C, No. 3**, pp. 132-140, Mar. 2014;

[8] S. Nagasawa and M. Hidaka, "Run-to-Run Yield Evaluation of Improved Nb 9-layer Advanced Process using Single Flux Quantum Shift Register Chip with 68,990 Josephson Junctions", *IOP Conf. Series: Journal of Physics: Conf. Series* **871,** 012065, 2017

[9] M. Tanaka, M.Kozaka, Y. Kita, et al., "Rapid Single-Flux-Quantum Circuits Fabricated Using 20-kA/cm$^2$ Nb/AlOx/Nb Process", *IEEE Trans. Appl. Supercond.*, **vol. 25, No. 3**, 1100304, June 2015.

[10] Rapid Single Flux Quantum (RSFQ) – Design Rules for Nb/Al$_2$O$_3$-Al/Nb-Process at Leibniz IPHT, Version 10.03.2017: RSFQ1H-1.6, available online: https://www.fluxonics.de/fluxonics-foundry/

[11] R. Monaco, R. Cristiano, L. Frunzio, and C. Nappi, "Investigation of low-temperature I-V curves of high -quality Nb/AI-AlOx/Nb Josephson junctions", *J. Appl. Phys.* **71 (4),** 1888-1892, 1991

[12] Y. Wu, L. Ying, G. Li, et al., "Film Stress Influence on Nb/Al-AlOx/Nb Josephson Junctions", *IEEE Trans. Appl. Supercond.*, **vol. 29, No. 5**, 1102105, August 2019.

[13] X. Zhang, et. al., Process "Control Monitoring for Fabrication Technology of Superconducting Integrated Circuits", submitted to ASC2020. Presentation ID: **Wk2EPo2A-03**

[14] X. P. Gao, et al., "Design and Verification of SFQ Cell Library for Superconducting LSI Digital Circuits", submitted to ASC2020. Presentation ID: **Wk1EPo3B-03**

[15] A.W. Kleinsasser, R.E. Miller, W.H. Mallison, Dependence of Critical Current Density on Oxygen Exposure in Nb-Al-Nb Tunnel Junctions, *IEEE Trans. Appl. Supercond.*, **5,** 26-30,1995